# Modified Dijkstra Algorithm with Invention Hierarchies Applied to a Conic Graph

**Ugochi A. Okengwu[1], Enoch O. Nwachukwu, Emmanuel N. Osegi[2]**

[1]Department of Computer Science, University of Port Harcourt, Port Harcourt, Nigeria
[2]Department of Information Technology, National Open University of Nigeria, Lagos, Nigeria
Email: Ugodepaker@yahoo.com, Enoch.nwachukwu@uniport.edu.ng, geeqwam@gmail.com

## Abstract

A modified version of the Dijkstra algorithm using an inventive contraction hierarchy is proposed. The algorithm considers a directed acyclic graph with a conical or semi-circular structure for which a pair of edges is chosen iteratively from multi-sources. The algorithm obtains minimum paths by using a comparison process. The comparison process follows a mathematical construction routine that considers a forward and backward check such that only paths with minimum lengths are selected. In addition, the algorithm automatically invents a new path by computing the absolute edge difference for the minimum edge pair and its succeeding neighbour in $O(n)$ time. The invented path is approximated to the hidden path using a fitness criterion. The proposed algorithm extends the multi-source multi-destination shortest-path problem to include those paths for which a path mining redirection from multi-sources to multi-destination is a minimum. The algorithm has been applied to a hospital locator path finding system and the results were quite satisfactory.

## 1. Introduction

Dijkstra algorithm is an iterative scheme that seeks to find the shortest path from a source node or vertex (position of exit) to a destination node (or target (position of entrance) address. The standard Dijkstra model algorithm [1] is a powerful yet simple approach to the short-path problem – a problem with several variants. However, the Dijkstra model can degrade in performance when faced with large data graphs or graphs with special edge weights such as negative or complex edges. Also, the topology and geometry of the graph might require some specific features in the Dijkstra model which usually is not available by default.

In order to effectively apply the Dijkstra model in modern applications, it is usual to modify the basic Dijkstra algorithm such that it becomes suitable in the intended application(s). In particular, the contraction hierarchies with pre-processing introduced by Geisberger [2], hold a lot of promises and have been widely applied in diverse problem in large road networks.

In this paper, we present a variant of a very simple but constructive approach to hierarchical routing – the Contraction Hierarchies (CH's). We assume the nodes of a weighted acyclic di-graph of the form G = (V, E) are la-belled v in order of ascending importance where v is multi-source i.e. v = 1, 2, 3 … for a range of target destinations say w1, w2, w3… propagated in a conical or semi-circular form. We form the adjacency matrix and construct a hierarchy by inventing or growing new nodes in the adjacency matrix.

An edge t1_t2 is invented by computing the absolute difference of two adjacent edge pair say s-t1 and s-t2, one of which is slightly smaller than the other and then removing the nodes from the network after a comparison operation with a boundary criterion. The invented path can be seen as an alternate path to a second (adjacent) target destination from the found short-path destination. For the best case we might seek to compare this path to a real connecting path and perform a fitness test.

This contraction procedure is done in such a way that the shortest-paths in the remaining overlay graph are preserved and that the invented node has been validated against a known fitness function.

The invention or contraction process here can be seen as a way to add all invented shortcuts to the edge set E, thus we obtain an inventive hierarchy (IH) for which we seek to find the minimum.

Here we emphasize edge weight contraction sufficiency for which we only invent when contraction might not be a realistically sufficient solution. Thus this work is an extension of CH and may even be modified further. Our approach may be viewed as a solution to the continuity test wherein we seek to evolve an edge hierarchy unconditionally with the hope that it would lead to better results. In this regard, a modified Dijkstra algorithm for an Extended Multi-source Multi-destination Shortest Problem with novel interesting contraction policy is proposed. We organize the paper as follows:

Section 2 reviews work areas relevant to this study describing a few customization attempts on the CH's; Section 3 briefly introduces the Dijkstra algorithm dissecting the key features of the algorithm. Section 4 introduces contraction hierarchies and expands this simple but useful idea to the possibly constructive invention hierarchy solution. Section 5 gives some experimental details and results for small sample dataset. Finally in Section 6 we give our conclusions and suggestions for further work.

## 2. Review of Relevant Works

Currently the work on modifications of the basic Dijkstra algorithm is an active one with key papers on the field discussing on its theory and applications. In this section we shall briefly discuss relevant works with a core focus on the CH's.

Huang [3] introduced a constraint function in their improved Dijkstra algorithm that depends on the size of the graph with node reduction priorities using open and closure principles. In their open lists, nodes that satisfy node reduction criterion are put into close list.

CH have been developed in [2] being a subset of the Highway-Hierarchies (HH) introduced in [4]. HH uses an expensive routine which necessitated the development of CH thus reducing the query time by two-fold however, with slight performance in node-ordering.

Rice [5] in their paper described a Klein Language Constrained Shortest Paths Problem (KLCSP) and developed modified Dijkstra implementations based on kleene language constructs for six types of node ordering metrics, thereby improving on the contraction hierarchy (CH) and solving the CH backward problem. The KLCSP can be seen as a "dynamically constrained shortest path query for large-scale datasets".

Hendrik [6] in his thesis work studied a multi-criteria path seeking scheme, and applied the concepts of A* and contraction with a> 1. In particular, they combined the CH and Label setting Algorithm [7] into a hybrid multi-criteria algorithm. In a multi-criteria problem, several instances are considered including the path-length, travel time, obstacles etc.

Funke in [8] described an extension of the CH's using a polynomial for the multicriteria problem with good heuristic solutions in the tri- and multi-criteria case.

Zeitz in [9], introduced the notion of weak CH's by removing the expensive witness search and with some slight reduction in performance and query time, their weak CH's showed promising results with a surprising better performance.

Schutz in [10] described a partition based Speed up technique for the Dijkstra algorithm. However, partitioning may result in an $O(n^3)$ complexity. Time dependent algorithms have been proposed in [11] and [12] with promising results.

Wang et al. [13] proposed a shortest-path algorithm for the multi-pair shortest path problem suitable for problems with a scattered Origin-Destination (OD) distribution.

However, the focus of all these researchers has been in the optimisation of existing paths and not necessarily the invention or discovery of new paths in the built-graph. In this paper, we seek to fill this gap by introducing an algorithmic idea that will not only encourage the discovery and invention of new paths based on a dynamic knowledge of hidden path existence but will as well encourage the search for sub-optimal short-path solutions for the invented paths. We shall also adopt a somewhat modified approach using a multi-pairs approach. For the purposes of this study we shall focus, on single-criteria scheme since this approach is fundamental to most short-path mining problems.

## 3. Basic Dijkstra Algorithm

Dijkstra algorithm [1] describes a technique to solve the single-source shortest path-problem. In particular, Dijkstra seek to reduce the cost of traversing through a set of paths called the edges which are formed by the interconnection of a pair of nodes. Mathematically shown in [16], Dijkstra computes the sum of important edges and ignores less important ones in somewhat greedy manner (see Table 1). To effectively implement this procedure Dijkstra emphasises a priority queue to perform key operations on the data graph effectively [4]:

- Insert key
- Delete Min
- Decrease key

**Table 1** Dijkstra Algorithm

*Function DIJKSTRA (G =< V, E, c, s >)*
*1: {The input to the algorithm is a directed graph G =< V; E >, weighted by the cost function $c : E \to Z^+$ ; we assume that there are no zero-cost edges}*
*2: **for** (i = 1 to n) **do***
*3: $d[i] = \infty$*
*4: **end for***
*5: d[s] = 0*
*6: Organize the vertices into a heap Q, based on their d values.*
*7: $S \leftarrow \emptyset$*
*8: **while** (Q $\neq \emptyset$) **do***
*9: u $\leftarrow$ EXTRACT-MIN(Q)*
*10: **for** (each edge of the form e = (u; v)) **do***
*11: RELAX(e)*
*12: **end for***
*13: S $\leftarrow$ S $\cup$ {u}*
*14: **end while***

## 4. Expanding Contraction Hierarchies (CH's)

### 4.1. Geisberger's CH

Contraction hierarchies introduced is a node ordering, and delete scheme that seeks to find out if a short-cut is necessary given a set of witnesses (witness paths). For every witness path mined, a comparison is made to the potential short-cut to determine if the short-cut is necessary (see Figure 1). If a short-cut is found, the in-between nodes are removed and preceding and succeeding nodes becomes a new formed edge. This is achieved in an additive fashion. Contraction Hierarchies are an extreme case of High-way Node Routing (HNR) where every node defines its own level of the hierarchy [14]. The pseudo-code for a CH algorithm is as shown in Table 2.

**Table 2** Geisberger's Contraction Hierarchies Algorithm

***Initializations:***
1. ***foreach** u    V ordered by < ascending **do***
2. ***foreach** (v, u)    E with u > v **do***
3. ***foreach** (u, w)    E with w > u **do***
4. ***if** (v, u, w) "may be" the only shortest path from v to w **then***
5.          E:= E    {(v, w)} (use weight w(v, w):= w(v, u) + w(u, w))

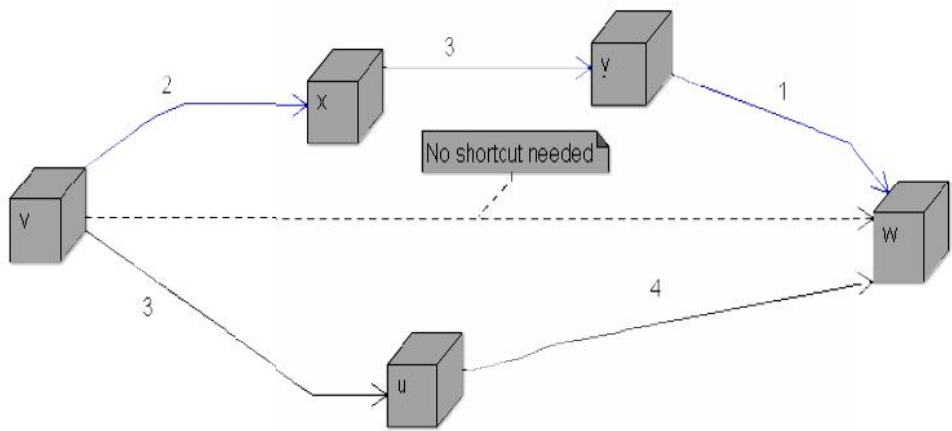

Figure 1. No short-cut because a witness P = (v, x, y, w) exists with w (v, u, w) >= w (P).

## 4.2. Graph Considerations and Contraction Policies

The nature of the graphs might likely influence our approach to contraction. Thus, a graph with a linear structure might contracted in forward only direction using solely amplified Dijkstra conditions along the paths eliminating long edges along path while traversal to target, whereas cyclic graphs will require a alternate elimination of nodes (or edges) along long paths and traversal through alternate paths (called short-cuts) to target. Graphs may also be directed (di-graphs) or undirected. For a conic, semi-circular or star di-graph this could mean traversal to potential (hypothetical) hidden targets while contracting adjacent boundary nodes or edges. Here, we are interested in weighted di-graphs with a somewhat conical or semi-circular structure.
These hypothetical targets may be mathematically constructed using known principles e.g. as in Pythagorean in geometry, triangle inequalities or complex-argument transformations, with the sole aim of inventing candidate short paths.

A contraction policy defines when a graph contraction procedure should proceed. It is paramount to the definition of a reasonable and realistic short-cut implementation paradigm.
Mathematically, short-cut policies follow a conditional inequality constraint which may be expressed generally as:

$$\dagger (x) \leq S_{allowable} . \qquad (4.1)$$

Where, (x) = stress developed anywhere in a component resource variable
$S_{allowable}$ = allowable strength of the material.
Here, the material refers to the graph under study and stress the path mining sequences involved.

### 4.3. Triangle Inequality

The triangle inequality is an age old mathematical principle that can be applied to a path mining algorithm's control logic to assure proper operation and coordination of a triangular structured algorithmic processing system. Stated in its simplest terms in [17] "no side of a triangle is greater in length than the sum of the other two sides nor less than the difference of the lengths of the other two sides mathematically restated here as:

$$|z1 + z2| \leq |z1| + |z2| . \qquad (4.2)$$

$$|z1 - z2| \geq |z1| - |z2| . \qquad (4.3)$$

In our system z1 and z2 represent edge pairs in an adjacency matrix for which we seek to find the absolute edge difference. In order to effectively utilize the triangle inequality in our short-path optimization problem and simplify our analysis, we introduce certain guiding axioms as provided below:

i. No two paths are equal
ii. We assume adjacent paths (edge or path-pairs) are close together, where we define close as angle of deviation between path-pairs that is constant and not greater than 30°
iii. We ignore the nth part if it invents an empty path

### 4.4. Hidden Paths

The notion of hidden paths stems from the fact that in a graph, not all paths are known. These unknown paths are called "hidden or secure" paths and may be discoverable through physical means by manually traversing through unknown zones from one edge destination to another nearby destination or mathematically by neighbourhood edge difference.
This scheme is captured in Figure 2. The Hidden path ideology can help capture the notion of feasibility in path mining and encourage the principle of inventive tactics in Graph theory.

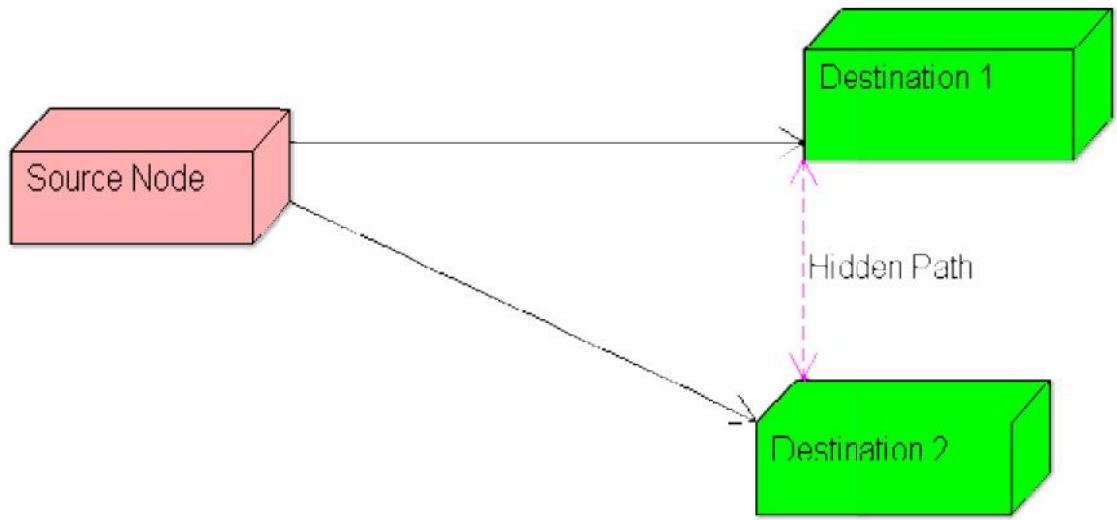

Figure 2. Hidden Path Concept

### 4.5. Invention CH Policy Formulation for Conic or Semi-circular Acyclic Graphs

A conic graph defines a group of network paths or arcs (small single-source conic graphs) for which there exists sub-paths bounded above (left) and below (right) by two or more major paths all paths emanating from a given case node. This is a convention and can be reversed to suit the developer's needs but the principle remains the same Note that above or below also may mean below or above so this principle is bi-directional.  Our system is a multi-source multi-destination graph. This situation is clearly depicted in Figure 3. For each source or origin node selected, we intelligently perform forward and backward comparisons in a forward-only search. In this way, a bi-directional search can be avoided and we can still solve the short-path problem.

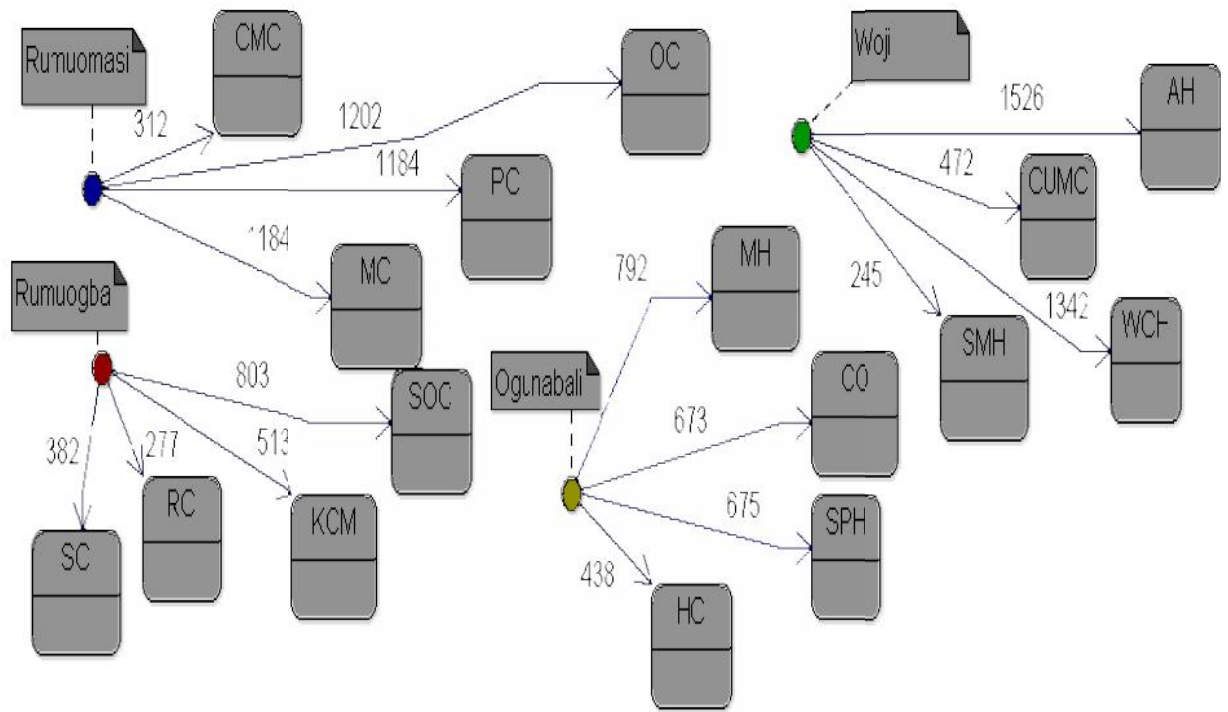

Figure 3 Conic multi-source multi-destination graph.

In order to understand while this graph has importance, consider a scenario where a patient proceeds to a destination hospital for which an alternate path exist to a neighbour hospital. Now, ideally, the patient follows the shortest path from the map, if it exists. Alternatively, the patient may seek a nearby hospital if current hospital does not offer requested service. In order to efficiently make this decision he might make approximate calculations on the best paths to take to nearest hospital. We simplify this calculation by using the concept absolute edge difference (AED) of adjacent neighbours. From experience, the patient would arrive at neighbour hospital after some trials. Such a path is coined "hidden-path" and should be fitted to AED. Ideally this is achieved using fitness (objective) measures or function. Constraints can be enforced mathematically by dynamically introducing maxima or minima point values (obstacles) for which the invented path must not proceed but this is beyond the scope of this paper. The dynamic implementation detail of the inventive contraction algorithm (ICH) is given in Table 3.

**Table 3** Inventive Algorithm

*Initializations:*

1. **foreach** $u \in V$ ordered by $<$ ascending **do**
2. **foreach** $(v, u) \in E$ with $u > v$ **do**
3. **foreach** $(v, w) \in E$ with $w > u$ **do**
4. **if** $(v, u)$ "may be" the only shortest path from $v$ to $w$ **then**
   a. $E := E \cup \{(v,u)\}$ (use weight $v(v,u)$)
   b. **else**
   c. $E := E \cup \{(v,w)\}$ (use weight $v(v,w)$)

5. $E := E \cup \{(v, u)\}$ (invent weight $w(v, k) := |v(v, k) - v(v, k+1)|$ )

     *where k=u or w. Note: k+1 ≻ v*

6.    *end*
7.    *end*
8.    *end*
9.    *end*

It is also possible to improve our algorithm search space further by introducing a dynamic orientation parameter Ө, for which alternate connecting edges can be constructed. This can greatly improve the short-path description graphic however this may introduce undesirable parameters into the system and make the algorithmic analysis awkward. Definition of a fitness function or objective function is a core requirement for achieving a conservative short-path minimization plan. In this regard, edge difference (AED) has been considered since this allows for norm estimates on error bounds and serve as an adaptive end-correction to invented paths.

## 5. Experiments

### 5.1. System Implementation and Environment

Our system is currently under development. We have performed some initial experiments on a small sample community-hospital route dataset extracted from the ARCGIS online data [18] for the Port-Harcourt Metropolis, Rivers State, Nigeria. The dataset has been entered into a MySQL database and analysed using a PHP debugging engine.

### 5.2. Node Ordering

Node ordering is a very vital part of the path mining process. Just as a good ordering scheme can assure the correctness of a query, a poorly constructed ordering scheme can reduce its performance and precision. Our node ordering scheme follows a predefined multi-pair sequence from which a build matrix is developed as shown in the build graph table (see Table 4). It uses a minimum select-adjacent search space locally within the graph. The target hospitals from their respective source locations (communities) are intuitively positioned column-wise and the corresponding communities row-wise. The intersection of the source communities and target hospitals give the edge values (distances) of the network. We assume that for each node, the next node is its neighbour node. In this way, our invention hierarchy formulation is assured. The destinations and sources are extensible and updatable since the MySQL offers a flexible, interactive and automated system.

Table 3 Simplified Transistion Matrix for the built graph

| Sources | | Destinations | | | | | | | |
|---|---|---|---|---|---|---|---|---|---|
| | | 0 | CMC | MC | PC | SC | PI | CU | OC | HC |
| | Offset | 0 | 1 | 2 | 3 | 4 | 5 | 6 | 7 | 8 |
| Rumuomasi | 0 | 0 | 312 | 771 | | | | | | |
| Runmuogba | 1 | | | | 374 | 382 | | | | |
| Woji | 2 | | | | | | 966 | 472 | | |
| Ogunabali | 3 | | | | | | | | 384 | 438 |

### 5.3. Application Instance

Query operation follows a unidirectional Dijkstra plan picking each pair of destination nodes for a selected source both in ascending order. We apply a Dijkstra search minimizing pairs of neighbour nodes in ascending order. Using the algorithm in Table 3.3 given in Section 4, we obtain new invented paths by selecting minimum edge for a given pair and succeeding neighbor. As an example, suppose:

v,u = 312, v,w = 771
then for (v:u, w)
minimum_Search = v(v,u)     = 312 {Since v,u < v,w
invented_hierarchy = |312-771| = 459

For a source node that possesses multiple connected edges, we can easily apply CH pre-processing to get a node reduct. For instance, if a shortcut is found for the edge values denoted v1, v2, v3 compared to a neighbor v11, v22, v33, we contract in the additive sense and obtain a simplified path represented as v1_v3 = v1+v2+v3 and v11_33 = v11 + v22 + v33.

We then invent a new interconnected edge abs (v1_v3 – v11_33) and use this as our basis for an extended shortest-path problem.

We can easily prove that the order of complexity of a CH reduction procedure is equivalent to that of a paired edge which is O (n).

### 6. Conclusion and Future Work

We have implemented a variant of contraction hierarchy using the notion of triangle inequality and approximate path projections between multi-paired bounded edges. Theoretical constructions of a variant class of contraction hierarchy that operates on a conic graph is implemented and proposed. Building on the underlying principles of variant contraction, further work will apply this approach in the Kleene Language Constraint Shortest Paths with empirical time complexity checks and expand the ideas developed here using well established contraction procedures. This should help improve on the dynamicity and performance of the graph processing algorithm. We would also suggest introducing the notion of hidden-bypass for which a cross-over may be necessary if a tentative destination becomes expensive and time-dependency is a core criterion. This can be done in such a way that a search can bypass its destination path by inventing a short-path to the witness neighbour if it is longer than the witness during its course, and hence finish its journey instead of returning back to its source. In particular the use of geometrical orientation parameter can help improve the accuracy of the estimation procedure and build in more dynamicity and realism in the path mining process since we are only using an approximation. This is analogous to transit node routing (TNR). Our next step will involve applying the developed algorithm on very large datasets using a variety of node-ordering metrics improved heuristics such as A* with time dependency, to improve the search direction towards finding the sources with the shortest invented paths